\documentclass{ws-procs975x65}

\begin{document}

\title{SCALAR FIELDS WITH BAROTROPIC EQUATION OF STATE: QUINTESSENCE VERSUS PHANTOM}

\author{OLGA SERGIJENKO$^*$ and BOHDAN NOVOSYADLYJ}

\address{Astronomical Observatory of Ivan Franko National University of Lviv,\\
Kyryla i Methodia str., 8, Lviv, 79005, Ukraine\\
$^*$E-mail: olka@astro.franko.lviv.ua}

\begin{abstract}
We constrain the parameters of dynamical dark energy in the form of a classical scalar field with barotropic equation of state jointly with other cosmological parameters using various combined datasets including the CMB power spectra from WMAP7, the baryon acoustic oscillations in the space distribution of galaxies from SDSS DR7 and WiggleZ, the light curves of SN Ia from 3 different compilations: SDSS (SALT2 and MLCS2k2 light curve fittings), SNLS3 and Union2.1. The considered class of models involves both quintessential and phantom subclasses. The analysis has shown that the phantom models are generally preferred by the observational data. We discuss the effect of allowing for non-zero masses of active neutrinos, non-zero curvature or non-zero contribution from the tensor mode of perturbations on the precision of dark energy parameters estimation. We also perform a forecast for the Planck mock data.
\end{abstract}


\bodymatter

\section*{}

Cosmological scalar fields are among the simplest and most promising candidates for dark energy. Up to now, many different Lagrangians and potentials have been studied. Here we consider the classical scalar field with barotropic equation of state. Such class of models involves both quintessential and phantom subclasses. We include into analysis the subclasses of models without peculiarities in the past \cite{Novosyadlyj2013}:
\begin{itemize}
 \item $w_0>-1$, $c_a^2>-1$;
 \item $w_0>-1$, $c_a^2<-1$;
 \item $w_0<-1$, $c_a^2<-1$, $c_a^2<w_0$.
\end{itemize}
We exclude the folowing subclasses of models, which can lead to $\rho_{tot}<0$ at some time in the past:
\begin{itemize}
 \item $w_0<-1$, $c_a^2>-1$;
 \item $w_0<-1$, $c_a^2<-1$, $c_a^2>w_0$.
\end{itemize}

\begin{figure}[htb]
\begin{center}
\psfig{file=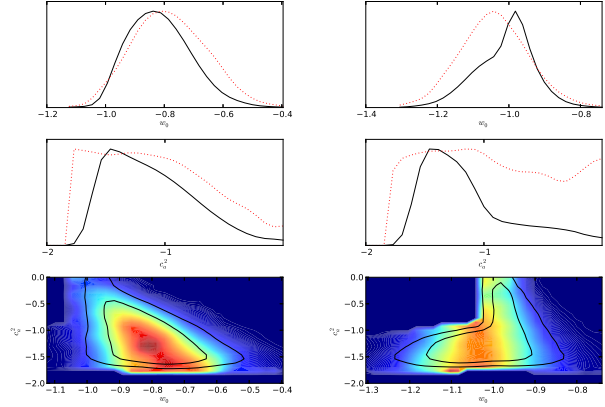,width=0.75\textwidth}
\end{center}
\caption{One-dimensional marginalized posteriors (solid lines) and mean likelihoods (dotted lines) for $w_0$ (top panels) and $c_a^2$ (middle panels). Left: WMAP7 {+} HST {+} BBN {+} BAO SDSS {+} SN SDSS MLCS2k2. Right: WMAP7 {+} HST {+} BBN {+} BAO SDSS {+} SN SDSS SALT2. Bottom: the corresponding two-dimensional mean likelihood distributions in the plane $c_a^2-w_0$. Solid lines show the $1\sigma$ and $2\sigma$ confidence contours.}
\label{fig}
\end{figure}

We have determined the best-fit values and confidence limits of the model parameters using the Markov chain Monte Carlo (MCMC) technique (implemented in the code CosmoMC)\cite{cosmomc} and the following data:
\begin{itemize}
 \item \textit{CMB temperature fluctuations and polarization angular power spectra} from the 7-year WMAP observations (WMAP7) 
\cite{wmap7_1,wmap7_2};
 \item \textit{Baryon acoustic oscillations} in the space distribution of galaxies from SDSS DR7 (BAO) \cite{bao_sdss};
 \item \textit{Hubble constant measurements} from HST (HST) \cite{hst};
 \item \textit{Big Bang Nucleosynthesis prior} on baryon abundance (BBN) \cite{bbn_1,bbn_2};
 \item \textit{supernovae Ia luminosity distance moduli} from SDSS compilation \cite{sn_sdss} with MLCS2k2 \cite{Jha2007} (SN SDSS MLCS2k2) and the SALT2 \cite{Guy2007} (SN SDSS SALT2) methods of light curve fitting.
\end{itemize}
The results for the combined datasets WMAP7 {+} HST {+} BBN {+} BAO SDSS {+} SN SDSS MLCS2k2 and WMAP7 {+} HST {+} BBN {+} BAO SDSS {+} SN SDSS SALT2 are presented in Fig. \ref{fig} and in Table \ref{tab}. We see that the dataset WMAP7 {+} HST {+} BBN {+} BAO SDSS {+} SN SDSS SALT2 prefers the phantom models of dark energy with barotropic equation of state, while the dataset WMAP7 {+} HST {+} BBN {+} BAO SDSS {+} SN SDSS MLCS2k2 gives preference to the models with $w_0>-1$, $c_a^2<-1$ (in accordance with conclusions of [\refcite{Novosyadlyj2013}]).

We have also used the newer data on SNe Ia distance moduli from
\begin{itemize}
 \item SNLS3 compilation (SNLS3) \cite{sn_snls} and
 \item Union2.1 compilation (Union2.1) \cite{sn_union}
\end{itemize}
together with data on BAO from the WiggleZ Dark Energy Survey (BAO WiggleZ) \cite{bao_wigglez}. The analysis of combined datasets WMAP7 {+} HST {+} BBN {+} BAO SDSS {+} BAO WiggleZ {+} SNLS3 and WMAP7 {+} HST {+} BBN {+} BAO SDSS {+} BAO WiggleZ {+} Union2.1 has shown that these data prefer the phantom models. Allowing for non-zero masses of active neutrinos, non-zero curvature or non-zero contribution from the tensor mode of perturbations does not change this conclusion. The obtained constraints on the massive active neutrino fraction of dark matter $f_{\nu}$, the curvature of 3-space $\Omega_k$ and the contribution from the tensor mode of perturbations $r$ are consistent with zero values of these parameters. The forecast made for the Planck mock data (generated using the code FuturCMB \cite{bluebook}) suggests that the models with $c_a^2>-0.75$ may be ruled out at $2\sigma$ confidence level by the Planck data.

\begin{table}[htb]
\tbl{The best-fit values, mean values and 2$\sigma$ marginalized confidence ranges for cosmological parameters determined by the MCMC technique using two observational datasets: WMAP7 {+} HST {+} BBN {+} BAO SDSS {+} SN SDSS MLCS2k2 and WMAP7 {+} HST {+} BBN {+} BAO SDSS {+} SN SDSS SALT2. The rescaled energy density of the component $X$ is denoted by $\omega_X \equiv \Omega_Xh^2$.}
{ \begin{tabular}{ccccc}
\toprule
Parameter&\multicolumn{2}{c}{MLCS2k2}&\multicolumn{2}{c}{SALT2}\\
&best-fit&2$\sigma$ c.l.&best-fit&2$\sigma$ c.l.\\
\colrule
$\Omega_{de}$& 0.702& 0.700$_{- 0.034}^{+ 0.031}$& 0.725& 0.725$_{- 0.030}^{+ 0.027}$\medskip\\
$w_0$&-0.758&-0.814$_{- 0.170}^{+ 0.228}$&-1.049&-1.010$_{- 0.171}^{+ 0.145}$\medskip\\
$c_a^2$&-1.295&-1.112$_{- 0.464}^{+ 0.672}$&-1.486&-1.139$_{- 0.423}^{+ 0.830}$\medskip\\
$10\omega_b$& 0.230& 0.227$_{- 0.011}^{+ 0.011}$& 0.224& 0.225$_{- 0.010}^{+ 0.010}$\medskip\\
$\omega_{cdm}$& 0.110& 0.110$_{- 0.009}^{+ 0.009}$& 0.114& 0.113$_{- 0.009}^{+ 0.009}$\medskip\\
$h$& 0.667& 0.665$_{- 0.028}^{+ 0.030}$& 0.704& 0.702$_{- 0.029}^{+ 0.029}$\medskip\\
$n_s$& 0.975& 0.974$_{- 0.026}^{+ 0.027}$& 0.966& 0.969$_{- 0.026}^{+ 0.026}$\medskip\\
$\log(10^{10}A_s)$& 3.075& 3.083$_{- 0.068}^{+ 0.071}$& 3.081& 3.086$_{- 0.066}^{+ 0.069}$\medskip\\
$\tau_{rei}$& 0.089& 0.090$_{- 0.024}^{+ 0.026}$& 0.080& 0.088$_{- 0.023}^{+ 0.025}$\medskip
\colrule
$-\log L$&\multicolumn{2}{c}{ 3857.113}&\multicolumn{2}{c}{ 3864.929}\\
\botrule
 \end{tabular}
}
\label{tab}
\end{table}

\bibliographystyle{ws-procs975x65}

\end{document}